# INDIRECT *ALLEE EFFECT*, BISTABILITY AND CHAOTIC OSCILLATIONS IN A PREDATOR-PREY DISCRETE MODEL OF LOGISTIC TYPE

**Ricardo López-Ruiz** [*]
**Danièle Fournier-Prunaret** [#]

[*] Department of Computer Science and BIFI,
Facultad de Ciencias-Edificio B,
Universidad de Zaragoza,
50009 - Zaragoza (Spain).

[#] Institut National des Sciences Appliquées,
Systèmes Dynamiques (SYD), L.E.S.I.A.,
Avenue de Rangueil, 31077 Toulouse Cedex (France).

## Abstract

A cubic discrete coupled logistic equation is proposed to model the predator-prey problem. The coupling depends on the population size of both species and on a positive constant $\lambda$, which could depend on the prey reproduction rate and on the predator hunting strategy. Different dynamical regimes are obtained when $\lambda$ is modified. For small $\lambda$, the species become extinct. For a bigger $\lambda$, the preys survive but the predators extinguish. Only when the prey population reaches a critical value then predators can coexist with preys. For increasing $\lambda$, a bistable regime appears where the populations apart of being stabilized in fixed quantities can present periodic, quasiperiodic and chaotic oscillations. Finally, bistability is lost and the system settles down in a steady state, or, for the biggest permitted $\lambda$, in an invariant curve. We also present the basins for the different regimes. The use of the critical curves lets us determine the influence of the zones with different number of first rank preimages in the bifurcation mechanisms of those basins.

# 1. PREDATOR-PREY MODELS: BENEFIT VS. DAMAGE IN TWO INTERACTING SPECIES

Life in an ecosystem is a complex system. Depending on the scale of observation one can try to understand the interaction among individuals of the same species or one can dissert about the relationship among different species in a higher scale. In the first case we are doing sociology of a particular species and in the second case we are studying ecology. In any case, the detailed analysis of each problem reveals new variables outside the scale of observation that are important to explain the dynamical behaviour of the system. Thus, for instance, the human society can not be understood as only a set of interacting human beings without the constraints imposed by the social super-structures such as institutions, public administration, enterprises, etc. [Situngkir, 2003], and by a similar argument, the evolution of rabbits and foxes populations is not understandable independently of an entangled system of interactions with other species in their respective environments [Murray, 2002].

A first approach to population dynamics distinguishes three main types of interaction between individuals or species, namely the predator-prey situation, competition or mutualism. Each one of them can be observed at both scales of observation but none of them can solely explain the richness of the living world. Still further, depending on the circumstances, the same individuals show a symbiotic behaviour in a moment, for instance the search for food, and a strong competition in other moments, such as the look for a couple with a procreation aim. Therefore, these interaction schemes are not found in a pure way if not confounded among them in many cases. Thus, the model of Lotka and Volterra for the predator-prey interaction between species is a not very realistic first-step approach [Lotka, 1925 ; Volterra, 1926]. Prey population may grow infinitely when there are no predators: any resource limitation in the prey number growth is not considered. Predators extinguish as well when there are no preys: they have no possibility to survive by consumption of other foods. If both species are present the rate of prey hunting by predators is proportional to prey density and the model shows an unnatural oscillation without asymptotic stability.

Numerous modifications of this model exist trying to make it more realistic.
A general predator-prey model with distributed delay in both, prey and predator, equations has been considered in [Krise & Choudhury, 2003]. In this study, a memory nonnegative bounded function reflects the influence of the past prey population values on the current dynamics. The integral term present in both equations represents a saturation limit in the prey dynamics and the prey's contribution to the predator growth in the first and second equations, respectively. The periodic structure appearing after the Hopf bifurcation is analyzed in the case of a weak kernel as memory function by the method of multiple scales.
A modified version of this model with distributed delays in both populations has been analyzed for weak and strong kernels [Dodd, 1997 ; Li *et al*., 2004]. They show that under certain conditions stable periodic solutions arising from Hopf bifurcations can exist in both cases.
In the same line of investigation, another possibility for a Lotka-Volterra predator-prey model is to consider discrete delays [Kuang, 1993]. One of such studies can be found in [Faria, 2001], where a Hopf bifurcation is proved to exist and it is described by using one of the delays as a system parameter. A similar work with two delays can be found in [Son *et al*.].
When more than a prey is allowed to the predator, a battery of models appears in function of the number of preys. As an example, we can mention an interacting one-predator-two preys model with harvesting of the predator at a constant rate. The periodic solutions arising from stationary states and their stability are studied as a function of the predator harvest rate [Kumar *et al*., 2002].

A further step consists in including a spatial dependence of the interacting populations. The individuals spread out in their environment as a result of an irregular motion. It results in a diffusion process which is modelled by reaction-diffusion equations. The Fisher-Kolmogoroff model for one species, where logistic evolution and diffusion are combined, is an example of this type of models [Fisher, 1937 ; Kolmogorov *et al.*, 1937]. The predator-prey model with discrete delays proposed in [Faria, 2001] was also studied with diffusion terms for both species. The Hopf bifurcation, and then the cyclic behaviour in the population evolution, continues to exist as in the original model. The stability of the periodic orbits emerging from this bifurcation in this type of systems is also determined in different papers [Faria & Huang, 2002].

In general it is difficult to obtain a model that accurately reflects the real behaviour in nature. The first goal of all these models is to obtain a regime where the out-of-phase periodic oscillation of the prey and predator populations is finally possible. Only in a punctual way, as it is discussed in [Murray, 2002] with the real lynx-hare data obtained from the number of pelts sold by the Hudson Bay Company, some author has suggested [Schaffer, 1984] that the predator-prey problem could show some evidence of a strange attractor. Let us remark at this point that the majority of the models here sketched are two-dimensional quadratic continuous systems where the irregular dynamics is not possible. Some ingredient expressed in an additional evolution equation would be missing in order to obtain such a chaotic behaviour. Let us also observe that the phenomenon of multi-stability is absent in the common predator-prey models and hence the species have no possibility to evolve to different final asymptotic states depending on their initial conditions.

In this paper, we propose a model issued from logistic maps where bistability, chaos and fractal behaviours can be obtained.

## 2. A NEW PREDATOR-PREY MODEL OF DISCRET LOGISTIC TYPE

### 2.1 The Logistic Model for One Isolated Species

Logistic evolution in time has been used to model one-species population dynamics [Verhulst, 1845 ; May, 1976]. If $x_n$ represents the population after $n$ generations, let us suppose this variable bounded in the range $0 < x_n < 1$. Initially the dynamics is controlled by the term $\mu x_n$ proportional to the current population $x_n$ and to the constant *growth rate* $\mu$. It is the *activation or expanding phase*. Resource limitations must saturate the exponential growth of this regime and should not allow the system to fall down in the overpopulation. The term $(1 - x_n)$ can give account of how far the system is from overcrowding. It produces an *inhibition or contracting phase* in the dynamics. Therefore, if we take the product of both last terms, the discrete logistic equation presents the form

$$x_{n+1} = \mu x_n (1 - x_n).$$

This model is the result of a simple approach for the evolution of an isolated species with limited affordable resources. Obviously the parameter $\mu$ is in the range $0 < \mu < 4$ in order to assure $0 < x_n < 1$. The continuous version of this model was originally introduced by Verhulst in the nineteenth century as an expression of the interplay between the Malthusian growth and the saturation imposed by the food limitation. The discrete version has been a subject of study in the last century as a tool to be applied to the most diverse phenomenology [Kaneko, 1983 ; Kendall & Fox, 1998] or as an object interesting to analyze by itself from a mathematical point of view [Collet & Eckmann, 1980 ; Mira, 1987].

The dynamical behaviour of the logistic equation when the growth rate is modified as follows:

(i) $0 < \mu < 1$: The growth rate is not big enough to stabilize the population. It will drop and the species will become extinct.
(ii) $1 < \mu < 3$: A drastic change is obtained when $\mu$ is greater than 1. The non vanishing equilibrium between the two competing forces, reproduction on one hand and resource limitation on the other, is now possible. The population reaches, independently of its initial conditions, a fixed value that is maintained in time.
(iii) $3 < \mu < 3.57$: A cascade of sudden changes provokes the oscillation of the population in cycles of period $2^n$, where $n$ increases from 1, when $\mu$ is close to 3, to infinity when $\mu$ is approaching the critical value 3.57. This is named the period-doubling cascade.
(iv) $3.57 < \mu < 3.82$: When the parameter moves, the system alternates between periodical behaviours with high periods on parameter interval windows and *chaotic regimes* for parameter values not located in intervals. The population can be not predictable although the system is deterministic. The chaotic regimes are observed for a given value of µ on sub-intervals of [0,1].
(v) $3.82 < \mu < 3.85$: The orbit of period 3 appears for $\mu = 3.82$ after a regime where unpredictable bursts, named *intermittences*, have become rarer until their disappearance in the three-periodic time signal. The existence of the period 3 orbit means, as the Sarkovskii theorem tells us, that all periods are possible for the population dynamics, although, in this case, they are not observable due to their instability. What is observed in this range is the period-doubling cascade $3*2^n$.
(vi) $3.85 < \mu < 4$: Chaotic behaviour with periodic windows is observed in this interval.
(vii) $\mu = 4$: The chaotic regime is obtained on the whole interval [0,1]. This specific regime produces dynamics, which looks like random. The dynamics has lost almost all its determinism and the population evolves as a random number generator.

Therefore, in addition to the rich birth process of its complex periodic orbit set, this system presents three remarkable dynamical behaviours: the period doubling route to chaos around the value $\mu \approx 3.57$ [Feigenbaum, 1978], the time signal complexification by intermittence in the neighbourhood of $\mu \approx 3.82$ [Pomeau & Manneville, 1980] and the random-like dynamics for $\mu = 4$.

## 2.2 The new Logistic Predator-Prey Model

Under a similar scheme of expansion/contraction, let us think in two interdependent species $(x_n, y_n)$ having a logistic evolution and interacting between them. One of the species $x_n$, the predator, is supposed to benefit by attacking the other species $y_n$, the prey, and on the contrary, this last species $y_n$ suffers damage in its population as a consequence of the attacks of the other species $x_n$. Obviously this interaction generates an asymmetrical coupling in the logistic-type evolution equations of this ecological system,

$$x_{n+1} = \mu_x(y_n)x_n(1-x_n),$$
$$y_{n+1} = \mu_y(x_n)y_n(1-y_n),$$

which presents a growth rate $\mu(z)$ varying with time as consequence of the predator-prey interaction. In fact, it depends on the population size of the *others* and of two positive constants $(\lambda_x, \lambda_y)$ associated in some way with the *reproduction rate* of both species:

$$\mu_x(z) = \lambda_x(3z+1),$$
$$\mu_y(z) = \lambda_y(-3z+4).$$

Hence, the dynamics of each particular species is a logistic map whose parameter $\mu_n$ is not fixed, $x_{n+1} = \mu_n x_n(1-x_n)$, but itself is forced to remain in the interval $[1,4]$. The existence of a nontrivial fixed point at each step $n$ ensures the nontrivial evolution of the system. The simplest election for this growth rates $\mu(z)$ is the product of the reproduction rate by an increasing or decreasing linear function of the other species size and expanding the whole interval $[1,4]$. Let us observe that if the populations are supposed to be uncoupled in some moment, i.e. $z=0$, then the fraction of the growth rates is $\frac{\mu_y}{\mu_x} = \frac{4\lambda_y}{\lambda_x}$. It seems to be realistic in a first approach to consider $\lambda_x = \lambda_y = \lambda$. This is an expression of a prey growth capacity four times stronger than that of the predator in the case that each species is living in isolation. When both species are put together in an ecosystem the *attack-defence* interaction takes place between them. It makes the parameter $\lambda$ to lose its pure reproductive meaning that it had in the case of isolation. It can now represent some kind of *mixed reproduction rate* containing also information on the mutual interaction between the species. This simplification facilitates our research in the sense that it is easier to explore with some detail the behaviour of the *final predator-prey model*,

$$\begin{aligned} x_{n+1} &= \lambda(3y_n+1)x_n(1-x_n), \\ y_{n+1} &= \lambda(-3x_n+4)y_n(1-y_n), \end{aligned} \quad (1)$$

as a function of the only parameter $\lambda$. This application can be represented by $T_\lambda : [0,1] \times [0,1] \to \Re^2$, $T_\lambda(x_n, y_n) = (x_{n+1}, y_{n+1})$, where $\lambda$ is a positive constant that the study has discovered to have sense in the range $0 < \lambda < 1.21$. In the following we shall write $T$ instead of $T_\lambda$ as the dependence on the parameter $\lambda$ is understood.

At this point we must remark that different choices of $\mu_n$ can produce other systems built under a similar insigth. For instance, the models, which we have denoted with the letters (a), (b) and (c), follow this insight of construction. Model (a) has been used to model the symbiotic interaction of two species [López-Ruiz & Fournier-Prunaret , 2004]. The dynamics and basin fractalization of the cubic model (b) have been studied in detail in [López-Ruiz & Fournier-Prunaret , 2003] and a similar study for the fifth degree model (c) is presented in [Fournier-Prunaret & López-Ruiz, 2003]. The on-off intermittence is another phenomenon that appears as a consequence of forcing $\mu_n$ to follow a random signal. The same idea can be applied in the continuous case. The study of the original logistic equation [Verhulst, 1845] by introducing time-dependent parameters in the population models can generate more realistic evolution results [Lakshmi, 2003 ; Leach & Andriopoulos, 2004]. Without diminishing the value of all these works, a battery of systems built with similar mechanisms is waiting to be explored and could be an interesting work to be performed in a next future.

In Sections 3 and 4, we study more accurately the model (1) from a dynamical point of view. In order to summarize, we explain first the dynamical behaviour of the coupled logistic system (1) of Lotka-Volterra type. When $\lambda$ is modified, we obtain:

(i) $0 < \lambda < 0.25$: The reproductive force of the preys is smaller than the combination of its natural death rate and the effect of predator attacks. Hence preys can not survive and the predators become also extinct.

(ii) $0.25 < \lambda < 0.4375$: Prey population can survive in a small quantity but predators do not have enough food to be self-sustained and become extinct. This phenomenon is some kind of *indirect Allee effect* for the case of two species, that is, if the prey population does not reach a certain threshold the predator population decreases in size until the extinction.

(iii) $0.4375 < \lambda < 1.051$: When the prey population exceeds the threshold $y^* \approx 0.43$, predator's attack strategy is efficient and the system settles down in an equilibrium where both populations are fixed in time. The increasing of $\lambda$ allows the coexistence of bigger populations: predator population increases faster than prey population up to reach a similar density around $x^* \approx y^* \approx 0.6$ where a new instability takes place.

(iv) $1.051 < \lambda < 1.0851$: For $\lambda = 1.051$, a stable period three cycle appears in the system. It coexists with the former fixed point. When $\lambda$ is increased, a period-doubling cascade takes place and generates successive cycles of higher periods $3 \cdot 2^n$. The system presents bistability. Depending on the initial conditions, both populations oscillate in a periodic orbit or, alternatively, settle down in the fixed point.

(v) $1.0851 < \lambda < 1.0997$: In this region, an aperiodic dynamics is possible. The period-doubling cascade has finally given birth to an order three cyclic chaotic band(s). The system can now present an irregular oscillation besides the stable equilibrium with final fixed populations.

(vi) $1.0997 < \lambda < 1.1758$: The basin of attraction of chaotic bands is absorbed by the one of the fixed point and the populations reach always the constant equilibrium, although in some cases, after a chaotic transient.

(vii) $1.1758 < \lambda < 1.211$: The system suffers a Hopf bifurcation giving rise to a stable invariant curve. The populations oscillate among a continuum of possible states located on the invariant curve.

(viii) $\lambda > 1.211$: The iterations are going outwards the square $[0,1] \times [0,1]$ and evolve towards infinity. The system 'crashes'. This critical value can be interpreted as some kind of catastrophe provoking the extinction of species.

Let us remark that there is no possible survival of predators without preys for a low reproduction rate. Only when the prey population reaches a threshold the predators can find an adequate hunting strategy to survive. This fact presents some similarity with the *Allee effect* for one species, that is, the decreasing in size of one species if this falls below a critical level. Another remarkable fact is the bistability between the fixed point and the high periodic orbits growing from the period doubling cascade of the period three cycle. And, finally, the possibility of a chaotic dynamics in the evolution of both species introduces an additional value in benefit of this model.

## 3. ATTRACTORS: NUMBER AND BIFURCATIONS

For the sake of clarity, firstly, we summarize the dynamical behaviour of model (1) when the reproduction rate $\lambda$ is inside the interval $0 < \lambda < 1.21091$. The different parameter regions where the mapping $T$ has stable attractors are given in the next table.

| INTERVAL | NUMBER OF ATTRACTORS | ATTRACTORS |
|---|---|---|
| $0 < \lambda < 0.25$ | 1 | $p_0$ |
| $0.25 < \lambda < 0.4375$ | 1 | $p_2$ |
| $0.4375 < \lambda < 1.051$ | 1 | $p_4$ |
| $1.051 < \lambda < 1.075$ | 2 | $p_4$, period three cycle |
| $1.075 < \lambda < 1.08511$ | 2 | $p_4$, period orbit of the doubling cascade $3 \cdot 2^n$ |
| $1.08511 < \lambda < 1.09967$ | 2 | $p_4$, period three chaotic bands |
| $1.09967 < \lambda < 1.17579$ | 1 | $p_4$ |
| $1.17579 < \lambda < 1.21091$ | 1 | invariant closed curve (or frequency lockings) |

The meaning of all these attractors is explained in the next sections.

### 3.1 Fixed Points and Closed Invariant Curve

In the range $0 < \lambda < 1.21091$, there exist stable attractors for each value of $\lambda$. The initial populations leading to each one of these possible asymptotic final states, that is, their basins of attraction, are studied in the next section.

The restriction of the map $T$ to the axes reduces to the logistic map $x_{n+1} = f(x_n)$, with $f(x) = \lambda x(1-x)$ for isolated predator population and $f(x) = 4\lambda x(1-x)$ in the case of the isolated preys. The fixed points are the solutions of $x_{n+1} = x_n$, which are $p_0, p_1, p_2$ on the axes and $p_3, p_4$ outside the axes:

$$p_0 = (0,0),$$
$$p_1 = \left(\frac{\lambda-1}{\lambda}, 0\right), p_2 = \left(0, \frac{4\lambda-1}{4\lambda}\right),$$
$$p_3 = \frac{1}{12p}\left\{(14p-3) + (4p^2+9)^{\frac{1}{2}}, 4(p-3) - 4(4p^2+9)^{\frac{1}{2}}\right\},$$
$$p_4 = \frac{1}{12p}\left\{(14p-3) - (4p^2+9)^{\frac{1}{2}}, 4(p-3) + 4(4p^2+9)^{\frac{1}{2}}\right\}.$$

For $0 < \lambda < 0.25$, $p_0$ is an attractive node. For all the rest of parameter values, $p_0$ is a repelling node. The points $(p_1, p_2)$ exist for every parameter value: $p_1$ is unstable for every value of $\lambda$ and $p_2$ is attractive only for $0.25 < \lambda < 7/16$. For $\lambda = 0.25$, $p_0$ and $p_2$ exchange stability in a transcritical bifurcation. When $\lambda = 7/16$, another transcritical bifurcation allows to exchange stability between $p_2$ and $p_4$. For $0.4375 < \lambda < 0.5194$, $p_4$ is an attractive node, and for $0.5194 < \lambda < 1.17579$ $p_4$ is a stable focus. For $\lambda = 1.17579$, $p_4$, which has coordinates $(0.68, 0.56)$, becomes unstable by a Neimark-Hopf bifurcation giving rise to a stable invariant curve (cf. Fig. 19). This invariant curve grows in size in the interval $1.17579 < \lambda < 1.21091$ and definitively loses stability for $\lambda = 1.21091$, when it suffers a crisis with the unstable heteroclinic connection formed by the period three cycle (Q'$_1$, Q'$_2$, Q'$_3$) $((0.48,0.49), (0.75,0.77), (0.74,0.36))$ (cf. Fig. 20). The point $p_3$ is unstable for every value of $\lambda$. It is located outside the square $[0,1] \times [0,1]$.

### 3.2 Period Three Cycle, Period Doubling Cascade and Chaotic Bands

A stable period three cycle appears after a saddle-node bifurcation for $\lambda = 1.051$. The coordinates of the points (Q$_1$, Q$_2$, Q$_3$) forming this cycle are $((0.35,0.38), (0.51,0.73), (0.84,0.50))$. The system presents bistability between this cycle and the fixed point $p_4$. The period three cycle doubles its period and the same takes place for the new orbits created. The parameters for which the birth of some of these cycles takes place are: $\lambda = 1.075$ for the $3 \cdot 2^1$ cycle, $\lambda = 1.0828$ for the $3 \cdot 2^2$ cycle, $\lambda = 1.0846$ for the $3 \cdot 2^3$ cycle, $\lambda = 1.085$ for the $3 \cdot 2^4$ cycle and so on. For $\lambda = 1.08511$, the period doubling cascade finishes for giving rise to a chaotic band with three sheets periodically visited (cf. Fig. 14). The crisis of this chaotic band with the boundary of the $p_4$ basin converts all this set of unstable orbits in a chaotic repellor (cf. Fig. 15). After this bifurcation, all the initial conditions with a finite asymptotic dynamics settle down on the steady state given by $p_4$ (cf. Fig. 16). Let us remark the limited existence on the parameter space of this additional periodical behaviour and the chain of bifurcations leading to its final chaotic state.

### 4. BASIN BEHAVIOUR

Let us see now how the different initial populations evolve towards its asymptotic stable state. It is worth noting that there are also phenomena in nature that are size dependent and the dynamics observed depends on the initial populations. In our mathematical representation, this is exactly the problem of considering the *basins* of the different attractors of model (1). For the sake of coherence, we consider the square $[0,1] \times [0,1]$ as the source of initial conditions having sense in our biological model, i.e., in the map $T$. Let us say at this point that basins constitute an interesting object of study by themselves. If a colour is given to the basin of each attractor, we obtain a coloured figure, which is a phase-plane visual representation of the asymptotic behaviour of the points of interest. The strong dependence on the parameters of this coloured figure generates a rich variety of complex patterns on the plane and gives rise to different types of basin fractalization. See, for instance, the work done by Gardini *et al.* [1994] and by López-Ruiz & Fournier-Prunaret [2003], and also by Tedeschini-Lalli [1995] in this direction, with coupled logistic mappings and with a predator-prey model, respectively. It is now our objective to analyze the parameter dependence of basins in model (1) by using the technique of critical curves.

## 4.1 Definitions and General Properties of Basins and Critical Curves

The set $D$ of initial conditions that converge towards an attractor at finite distance when the number of iterations of $T$ tends towards infinity is the basin of the attracting set at finite distance. When only one attractor exists at finite distance, $D$ is the basin of this attractor. When several attractors at finite distance exist, $D$ is the union of the basins of each attractor. The set $D$ is invariant under backward iteration $T^{-1}$ but not necessarily invariant by $T$: $T^{-1}(D) = D$ and $T(D) \subseteq D$. A basin may be connected or non-connected. A connected basin may be simply connected or multiply connected, which means connected with holes. A non-connected basin consists of a finite or infinite number of connected components, which may be simply or multiply connected. The closure of $D$ also includes the points of the boundary $\partial D$, whose sequences of images are also bounded and lay on the boundary itself. If we consider the points at infinite distance as an attractor, its basin $D_\infty$ is the complement of the closure of $D$. When $D$ is multiply connected, $D_\infty$ is non-connected, the holes (called lakes) of $D$ being the non-connected parts (islands) of $D_\infty$. Inversely, non-connected parts (islands) of $D$ are holes of $D_\infty$ [Mira *et al.*, 1996].

In Sec. 3, we explained that the map (1) may possess one or two attractors at a finite distance. The points at infinity constitute the third attractor of $T$. Thus, if a different colour for each different basin is given we obtain a coloured pattern in the square $[0,1] \times [0,1]$ with a maximum of two colours. In the present case, the phenomena of finite basins disappearance have their origin in the competition between the attractor at infinity (whose basin is $D_\infty$) and the attractors at finite distance (whose basin is $D$). When a bifurcation of $D$ takes place, some important changes appear in the coloured figure, and, although the dynamical causes cannot be clear, the coloured pattern becomes an important visual tool to analyze the behaviour of basins.

Critical curves are an important tool used to study basin bifurcations. They were introduced by Mira in 1964 (see [Mira *et al.*, 1996] for further details). The map $T$ is said to be noninvertible if there exist points in state space that do not have a unique rank-one preimage under the map. Thus the state space is divided into regions, named $Z_i$, in which points have $i$ rank-one preimages under $T$. These regions are separated by the so called critical $LC$ curves, which are the images of the $LC_{-1}$ curves: $LC = T(LC_{-1})$. If the map $T$ is continuous and differentiable, the $LC_{-1}$ curve is the locus of points where the determinant of the Jacobean matrix of $T$ vanishes. When initial conditions are chosen to both sides of $LC$ curve, the rank-one preimages appear or disappear in pairs. (See also the glossary for different technical terms used along this work).

## 4.2 Critical Curves and the $Z_i$ Regions of T

The map $T$ defined in (1) is noninvertible. It has a non-unique inverse. $LC_{-1}$ is the curve verifying $|DT(x, y)| = 0$, where $DT(x, y)$ is the Jacobean matrix of $T$. It is formed by the points $(x, y)$ that satisfy the equation:

$$27x^2y^2 + 3x^2y - 57xy^2 - 6x^2 + 24y^2 + 2xy + 11x - 4y - 4 = 0. \qquad (2)$$

Hence, $LC_{-1}$ is independent of $\lambda$ parameter and is quadratic in $x$ and $y$. It can be observed that $LC_{-1}$ is a curve of four branches, with two horizontal and two vertical asymptotes. The branches $LC_{-1}^{(11)}$ and $LC_{-1}^{(12)}$ have as horizontal asymptote the line $y = 0.419$ and the vertical asymptote in $x = 0.581$. The other two branches, $LC_{-1}^{(21)}$ and $LC_{-1}^{(22)}$, have the horizontal asymptote in $y = -0.581$ and the vertical one is the line $x = 1.530$. It follows that the critical curve of rank-1, $LC^{(ij)} = T(LC_{-1}^{(ij)})$, $i = 1,2$, $j = 1,2$, consists of four branches. The shape of $LC$ and $LC_{-1}$ is shown in Figs. 1-2. $LC$ depends on $\lambda$ and separates the plane into three regions that are locus of points having 1, 3 or 5 distinct preimages of rank-1. They are denoted by $Z_i$, $i = 1,3,5$, respectively (Figure 3). Observe that the set of those points with three preimages of rank-1, i.e. $Z_3$, is not connected and it is formed by five disconnected zones in the plane.

Let us observe that the four branched $LC$-curve divides the diagonal in five intervals for each $\lambda$. It is straightforward to verify the number of preimages of each one of the $Z_i$ zones by directly calculating this number for different points located in each one of the five intervals on the diagonal. A detailed calculation of a similar problem has been performed in [López-Ruiz & Fournier-Prunaret, 2003]. Thus, the point $(0,0)$ has five preimages, which are $\{(0,0),(0,1),(1,0),(1,1),(1.33,-0.33)\}$, independently of $\lambda$. For $\lambda = 0.5$, the point $(-1,-1)$ has three preimages which are $\{(-1.68,-0.18),(-0.30,1.30),(1.18,2.69)\}$, the point $(0.3,0.3)$ presents also three preimages which are $\{(0.24,0.76),(0.53,0.47),(1.56,-0.56)\}$, the point $(1,1)$ has an only preimage which is $\{(1.8,-0.8)\}$ and the point $(3,3)$ presents three preimages which are $\{(1.5,-3),(2.14,-1.14),(4,-0.50)\}$. When $\lambda$ is modified the pattern of the $Z_i$ zones is maintained. According to the nomenclature established in Mira *et al.* [1996], the map (1) is of type $Z_3 - Z_5 \succ Z_3 - Z_1 \prec Z_3$.

**4.3 Types of Regimes in T and its Basins**

Depending on $\lambda$, three different types of patterns are obtained in the square $[0,1] \times [0,1]$. We proceed to present them and to explain the role played by critical curves in the bifurcations of basins.

*4.3.1 Extinction of Species:* $0 < \lambda < 0.25$

In this regime, any given initial population evolves towards the extinction. The reproduction rate is too small to allow the surviving of the preys, and then, the predators are also extinguished. The iteration of all the initial conditions tends to zero, and a basin of only one colour is obtained (Fig. 4).

*4.3.2 Survival of Preys and Death of Predators:* $0.25 < \lambda < 0.4375$

The reproduction rate is now enough to maintain a small prey population. The strategy developed by the predators has no success and they become extinct. One colour pattern is also obtained in this regime (Fig. 5).

*4.3.3 Stable Coexistence of Preys and Predators:* $0.4375 < \lambda < 1.051$

When the critical value of prey population is reached the indirect *Allee effect* disappears and the predators can survive by hunting the preys (Fig. 6-7-8). When λ<1, the basin of the attractor is the whole square $[0,1]\times[0,1]$. When λ>1, small parts of the square $[0,1]\times[0,1]$ form part of the basin of infinity, and if the initial conditions are taken inside them a new kind of extinction is obtained. The cause is not the lack of resources if not some internal catastrophe of the system. This bifurcation is due to the crossing of $LC^{(11)}$ through corners (0,1) and (1,0) (cf. Fig.6). This crossing creates white parts inside $Z_3$ ($H_1$ and $H_2$) and their preimages of any order, called bays, are located on the border of the square $[0,1]\times[0,1]$ (cf. Fig. 6-7). See on Fig. 6 the rank-1 preimages $H_2^{-1}$ and $H_1^{-1}$. New crossings of the bays through LC curves give rise again to new bays on the borders of the bays and this phenomenon changes their shapes (cf. Fig. 7-8).

### *4.3.4 Stable Coexistence, Periodic Oscillations or Chaos:* $1.051 < \lambda < 1.0997$

The basin for $\lambda = 1.051$ suffers a sudden change. The period three cycle appears and a ball of initial conditions are attracted towards it. The coexistence between both species can reach a stable value or can oscillate with period three. The basin presents two colours (Fig. 9-14) : the red one corresponds to the basin of the fixed point and the yellow one to the basin of the period 3 orbit. This bicoloured pattern remains even after the period doubling cascade of the three cycle and its posterior chaotization, with some changes that we are going to explain.

When λ=1.058, there is a tangency between $LC^{(21)}$ and the yellow basin at the point $M_1$, the yellow basin has now a part inside $Z_5$ and touches the boundary of $Z_3$ area (cf. Fig. 10); it gives rise to a tangency between two non-connected parts, which are their preimages, located at $M_1^{-1}$, on $LC_1^{(21)}$ (see [Mira & al., 1994]). Such a bifurcation creates connections between non-connected parts of the basin, which changes its shape (cf. Fig. 11).

Another bifurcation corresponds to the appearance of holes inside the basin and can be explained as follows (cf. Fig. 12) : the $LC^{(11)}$ curve becomes tangent to the basin boundary, which creates a hole $H_0$ inside $Z_3$ area; the preimages of any rank of $H_0$ give rise to holes inside the basin, which becomes multiply connected. For instance, $H_0^{-1}$ is a first rank preimage of $H_0$, which did not exist before the tangency.

Moreover, the crossing of LC curves through yellow parts implies that new parts are created inside a $Z_i$ area, i=3,5, and it gives rise to more preimages of any order, in yellow colour, which accumulate along the basin boundary.

The last bifurcation in this parameter interval corresponds to a tangency between the order 3 chaotic cyclic attractor and its basin boundary (cf. Fig. 14a-b). This contact bifurcation gives rise to the chaotic attractor disappearance. There remains only a single stable state, which is the fixed point $P_4$. After the contact bifurcation, there exists a chaotic transient for iterated sequences issued from initial conditions chosen inside the basin of the previous chaotic attractor before reaching the stable fixed point (cf. Fig. 15).

### *4.3.5 Stable Coexistence:* $1.0997 < \lambda < 1.1758$

The fixed point is now the only possible final state. The basin of one colour is recovered (cf. Fig. 16-18). A new bifurcation occurs, leading to the opening of holes, which become bays (cf. Fig. 16-17). This bifurcation is of the same kind than the bifurcation of Fig. 10 : tangency between $LC^{(11)}$ curve and red basin gives rise to a connected – non connected parts bifurcation; bays open (see for instance $LC_{-1}^{(11)}$ curve on Fig. 16).

New holes appear again inside the basin by crossing of LC curves through the basin boundary, holes open to give bays and these bifurcations give rise to a fractal boundary of the red basin (see [Mira & al., 1994] for more details).

*4.3.6 Quasiperiodic Oscillation and Catastrophe of Species,* $1.1758 < \lambda < 1.2109$

The Hopf bifurcation for $\lambda = 1.1758$ gives rise to an invariant curve (cf. Fig. 19-20). This is the expression of a quasiperiodic dynamics. Then the stable state oscillates between a quasi-periodic state and periodic states of different orders, which corresponds to frequency lockings. Then, a collision with an unstable heteroclinic connection formed by the period three cycle $(Q'_1, Q'_2, Q'_3)$ makes the stable state disappear. Then any initial condition gives rise to an extinction of both species through an internal catastrophe.

## 5. CONCLUSIONS

Three main types of interaction among species, namely predator-prey situation, competition or mutualism among species, are usually taken as the brick ingredient for understanding how their populations evolve in an ecosystem. To our knowledge, independently of the kind of interaction, the evolution models that are found in the literature are stated with a quadratic dependence on the size populations. In this work, we have studied a cubic two-dimensional coupled logistic equation as a discrete model to explain the evolution of a predator-prey system. The coupling between both species is population-size dependent and is controlled by a positive constant $\lambda$ which could be related with the *reproduction rate* of preys and with the success of the *hunting strategy* of the predators. Depending on $\lambda$, the system can reach extinction for small $\lambda$, stable coexistence, and periodic or quasiperiodic oscillation of predator-prey populations. These dynamical regimes are commonly found in the majority of the predator-prey models. It is not the case for the indirect *Allee effect*, the *period doubling cascade* of a period three orbit and the final *chaotic state* that are present in this model. The fact that one of the species needs a critical level of the other population for its possible existence is a kind of indirect Allee effect. It appears in a natural way in this model. More surprising is the phenomenon of *bistability* between the orbits deriving from the period doubling cascade and the synchronized state. In this regime, the initial conditions determine the different final behaviour. The system can also reach a chaotic state as a new ingredient that enriches the dynamical possibilities of this model.

Different complex colour patterns on the plane have been obtained when the reproduction rate is modified. Critical curves have been used in order to understand the basin bifurcations for this map of $Z_3 - Z_5 \succ Z_3 - Z_1 \prec Z_3$ type. A detailed study of the different fractalization mechanisms for the whole range of $\lambda$ parameter for a similar coupled logistic equation was performed in [López-Ruiz & Fournier-Prunaret, 2003]. Finally, let us remark the diversity of the complex patterns produced on the plane and the different dynamical behaviours of this model.

**Acknowledgements**

We would like to thank Dr. Taha for his helpful comments. R. L.-R. wishes also to thank the *Systèmes Dynamiques* Group at INSA (Toulouse) for its kind hospitality, and M. Hernández-Pardo for very useful comments. This work was supported by the Spanish MCYT project HF2002-0076 and French research project EGIDE-PICASSO 05125VC.

# GLOSSARY

INVARIANT: A subset of the plane is invariant under the iteration of a map if this subset is mapped exactly onto itself.

ATTRACTING: An invariant subset of the plane is attracting if it has a neighbourhood every point of which tends asymptotically to that subset or arrives there in a finite number of iterations.

CHAOTIC AREA: An invariant subset that exhibits chaotic dynamics. A typical trajectory fills this area densely.

CHAOTIC ATTRACTOR: A chaotic area, which is attracting.

BASIN: The basin of attraction of an attracting set is the set of all points, which converge towards the attracting set.

IMMEDIATE BASIN: The largest connected part of a basin containing the attracting set.

ISLAND: Non-connected region of a basin, which does not contain the attracting set.

LAKE: Hole of a multiply connected basin. Such a hole can be an island of the basin of another attracting set.

HEADLAND: Connected component of a basin bounded by a segment of a critical curve and a segment of the immediate basin boundary of another attracting set, the preimages of which are islands.

BAY: Region bounded by a segment of a critical curve and a segment of the basin boundary, the successive images of which generate holes in this basin, which becomes multiply connected.

CONTACT BIFURCATION: Bifurcation involving the contact between the boundaries of different regions. For instance, the contact between the boundary of a chaotic attractor and the boundary of its basin of attraction or the contact between a basin boundary and a critical curve *LC* are examples of this kind of bifurcation.

# Figure Captions

**Fig. 1:** $LC_{-1}$ curves.
**Fig. 2 :** LC curves.
**Fig. 3 :** $Z_i$ areas.
**Fig. 4 :** The only stable state $P_0$ corresponds to the extinction of both species.
**Fig. 5 :** The parameter value increases, it permits the existence of preys only.
**Fig. 6 :** Bays appear along the basin.
**Fig. 7 :** The only stable state is the fixed point $P_4$.
**Fig. 8 :** Same as Fig. 7 with LC curves.
**Fig. 9 :** Basins of two distinct attractors, $P_4$ and period 3 orbit ($Q_1$, $Q_2$, $Q_3$).
**Fig. 10 :** Bifurcation non-connected parts – connected parts concerning the yellow basin.
**Fig. 11 :** Basins of two distinct attractors, $P_4$ and period 3 orbit ($Q_1$, $Q_2$, $Q_3$).
**Fig. 12 :** Holes appearance inside the basin.
**Fig. 13 a-b :** Order 3 cyclic chaotic attractor and enlargement.
**Fig. 14 a-b :** Contact bifurcation leading to order 3 attractor disappearance.
**Fig. 15 :** Chaotic transient leading to $P_4$.
**Fig. 16 :** Bifurcation connected parts – non connected parts.
**Fig. 17 :** Holes are open, they give bays.
**Fig. 18 :** Fractalization of basin.
**Fig. 19 :** A Neïmark-Hopf bifurcation gives rise to the invariant closed curve C.
**Fig. 20 :** Frequency locking before disappearance of C by contact bifurcation with its basin.

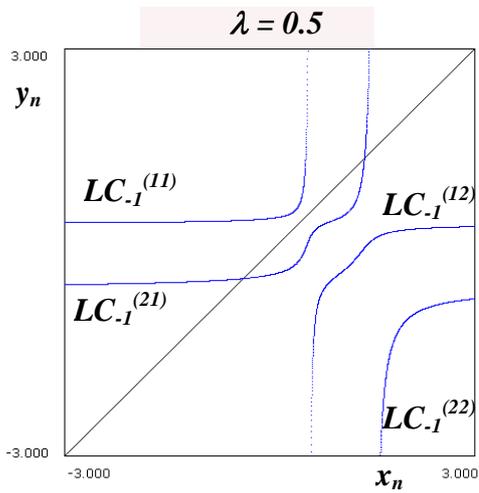
Figure 1

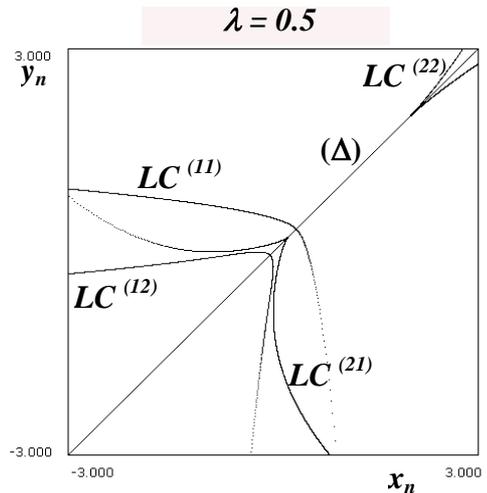
Figure 2

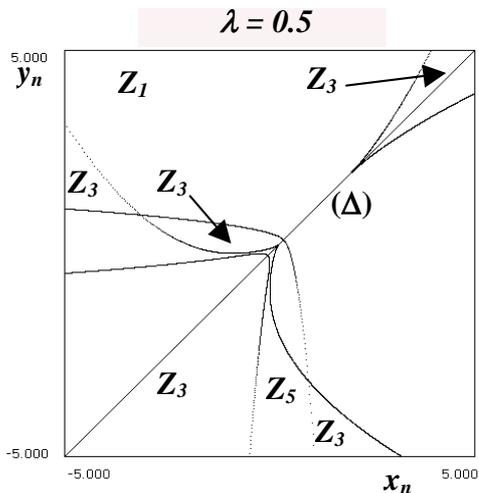
Figure 3 : $Z_i$ areas

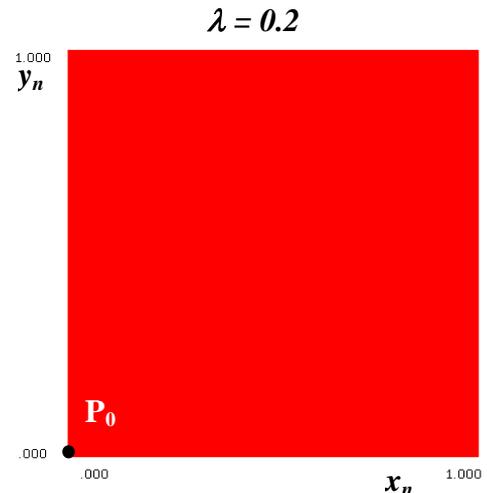
Figure 4

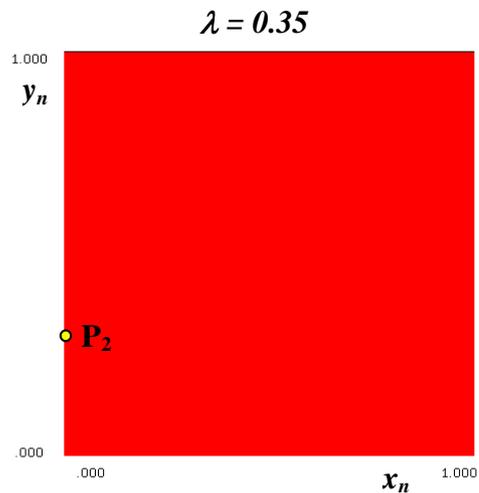
Figure 5

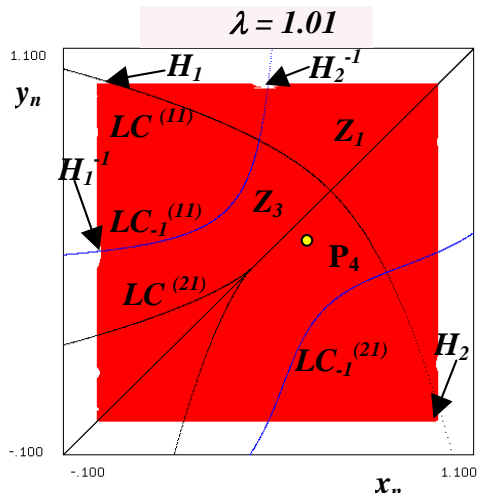

Figure 6

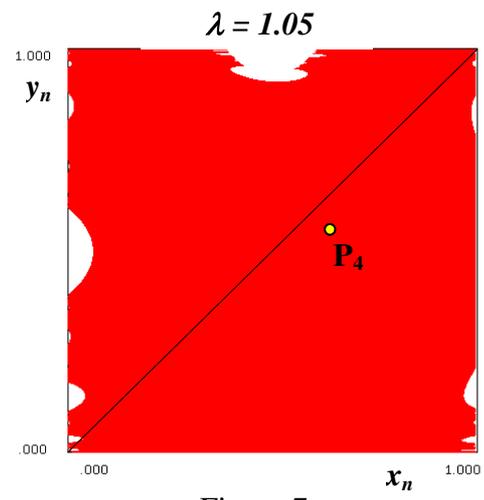

Figure 7

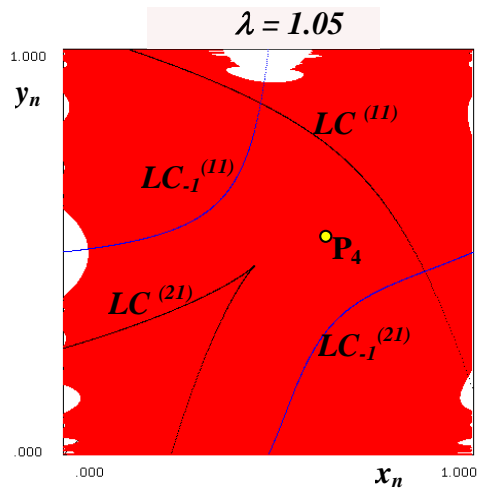

Figure 8

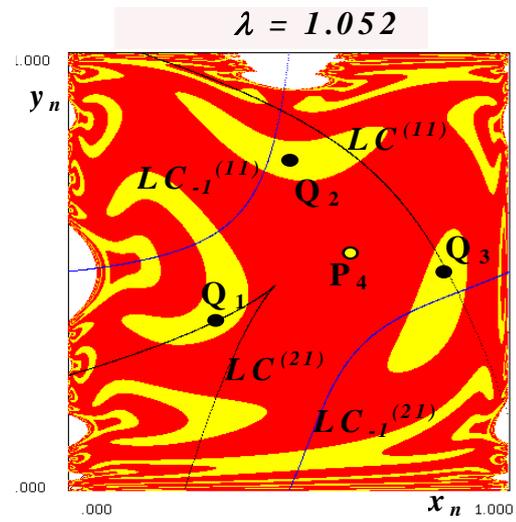

Figure 9

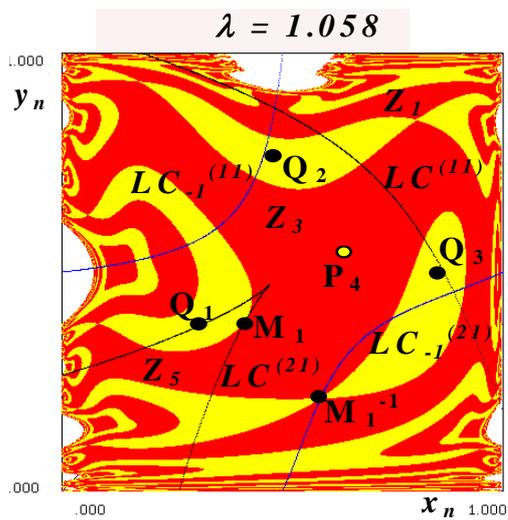

Figure 10

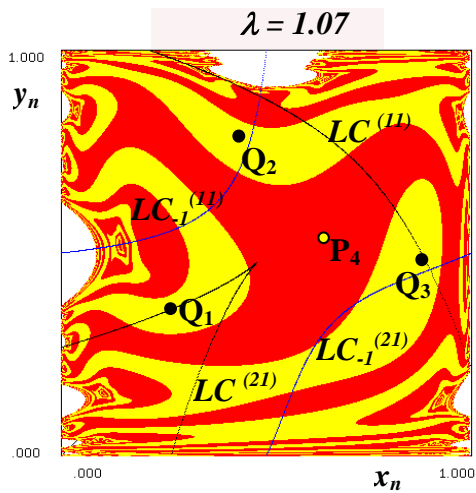
Figure 11

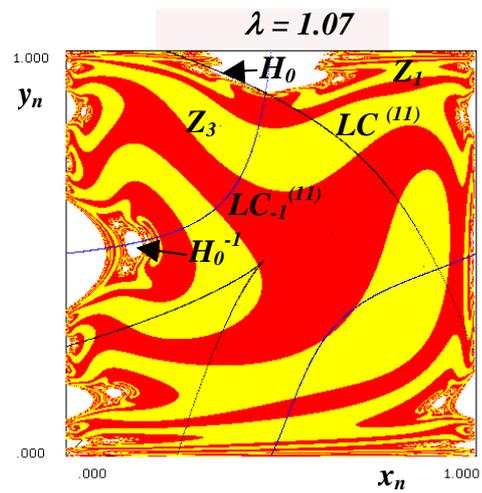
Figure 12

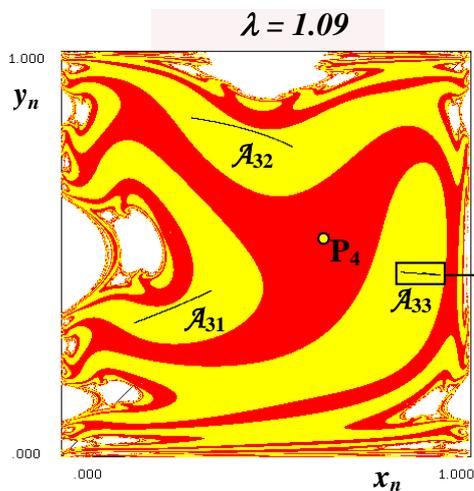
Figure 13a

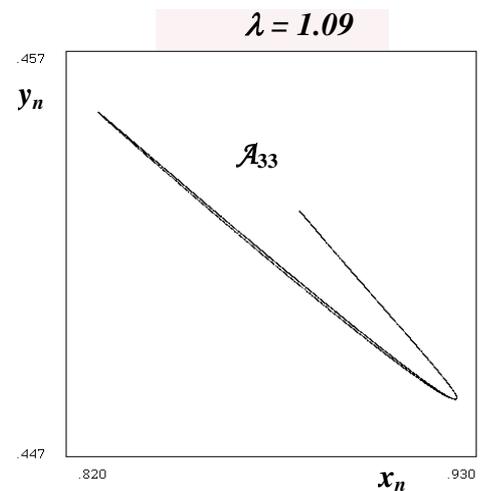
Figure 13b

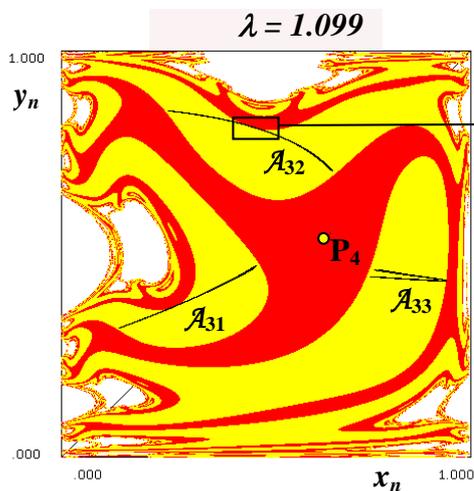
Figure 14a

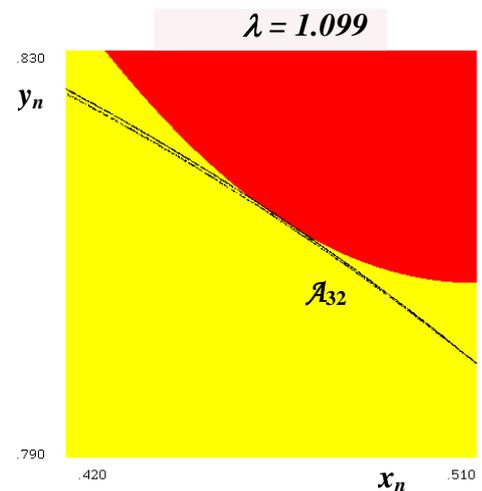
Figure 14b

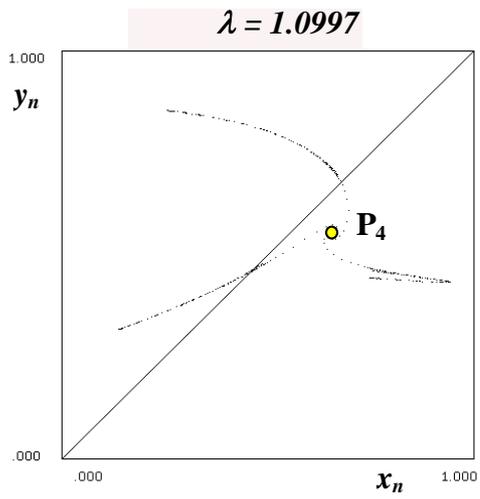

Figure 15

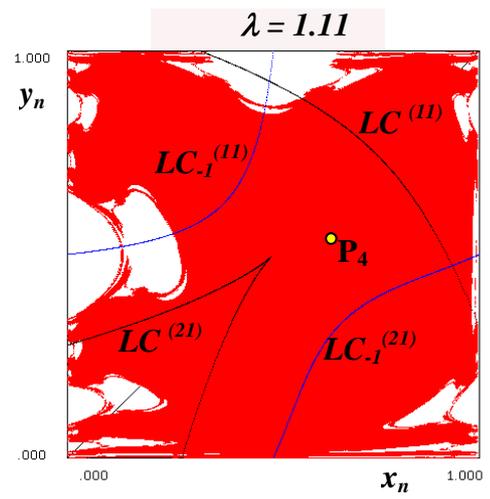

Figure 16

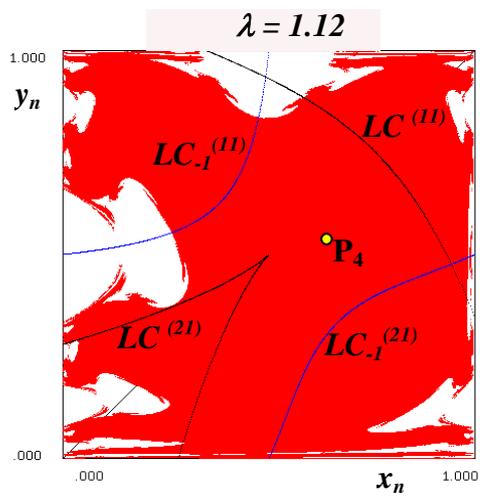

Figure 17

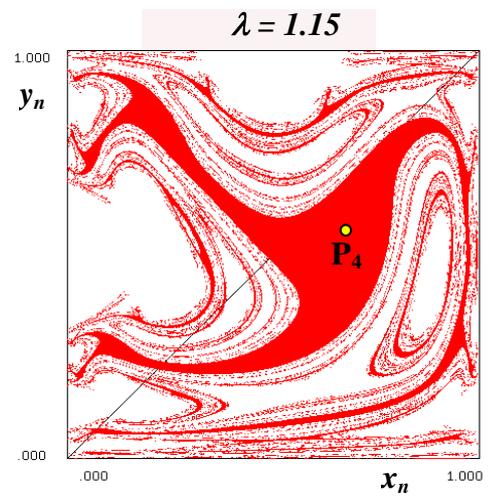

Figure 18

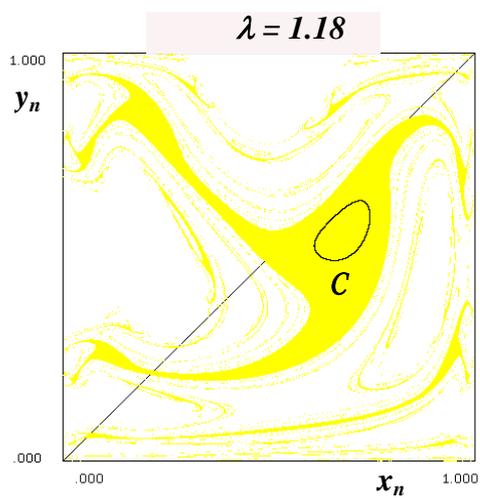

Figure 19

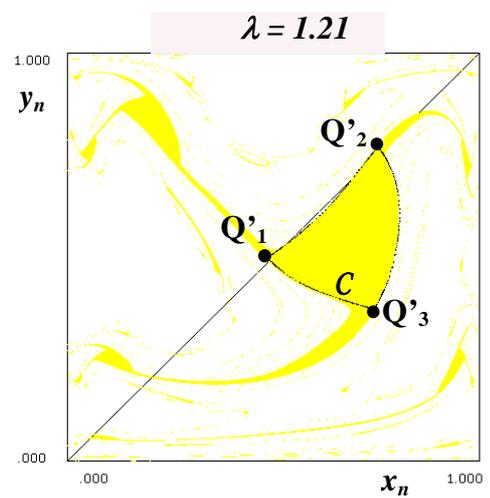

Figure 20